\documentclass[10pt,aps,pra,twocolumn,superscriptaddress]{revtex4-2}
\usepackage[T1]{fontenc}
\usepackage{amsmath, amssymb}
\usepackage[caption=false]{subfig} 
\usepackage{graphicx}
\usepackage{orcidlink}
\usepackage[stretch=10]{microtype}
\usepackage{float}
\usepackage[group-separator={}]{siunitx}
\usepackage{soul}
\usepackage{xcolor}
\usepackage{color}  
\usepackage{todonotes}
\usepackage{braket}
\usepackage[normalem]{ulem}
\usepackage{hyperref}
\hypersetup{colorlinks,linkcolor={red!50!black},citecolor={blue!60!black},urlcolor={blue!70!black}}

\newcommand{\grasp}{{\sc grasp2018}}

\newcommand{\invsec}{\unit{\per\second}}

\begin{document}

\author{Joseph S. Andrews\orcidlink{0009-0005-4535-8342}}
\affiliation{Theoretisch-Physikalisches Institut, Friedrich-Schiller-Universit\"at Jena, D-07743 Jena, Germany}
\affiliation{Helmholtz-Institut Mainz, D-55099 Mainz, Germany}
\affiliation{GSI Helmholtzzentrum f\"ur Schwerionenforschung GmbH, D-64291 Darmstadt, Germany}

\author{Andrey I. Bondarev\orcidlink{0000-0002-5880-4779}}
\affiliation{GSI Helmholtzzentrum f\"ur Schwerionenforschung GmbH, D-64291 Darmstadt, Germany}
\affiliation{Helmholtz-Institut Jena, D-07743 Jena, Germany}

\author{Per J\"onsson\orcidlink{0000-0001-6818-9637}}
\affiliation{Department of Materials Science and Applied Mathematics, Malm\"o University, S-205 06 Malm\"o, Sweden}

\author{\\Jon Grumer\orcidlink{0000-0002-6224-3492}}
\affiliation{Theoretical Astrophysics, Department of Physics and Astronomy, Uppsala University, Box 516, S-751 20 Uppsala, Sweden}

\author{Sebastian Raeder\orcidlink{0000-0003-0505-1440}}
\affiliation{Helmholtz-Institut Mainz, D-55099 Mainz, Germany}
\affiliation{GSI Helmholtzzentrum f\"ur Schwerionenforschung GmbH, D-64291 Darmstadt, Germany}

\author{Stephan Fritzsche\orcidlink{0000-0003-3101-2824}}
\affiliation{Theoretisch-Physikalisches Institut, Friedrich-Schiller-Universit\"at Jena, D-07743 Jena, Germany}
\affiliation{GSI Helmholtzzentrum f\"ur Schwerionenforschung GmbH, D-64291 Darmstadt, Germany}
\affiliation{Helmholtz-Institut Jena, D-07743 Jena, Germany}

\author{Jacek~Bieroń\orcidlink{0000-0002-0063-4631}}  
\affiliation{Instytut Fizyki Teoretycznej, Uniwersytet Jagielloński, 30-348 Kraków, Poland}

\title{\textit{Ab initio} multiconfigurational calculations of experimentally significant energy levels and transition rates in Lr~I $\left( Z=103 \right)$}

\date{\today}

\begin{abstract}
    Large-scale multiconfigurational calculations are conducted on experimentally significant transitions in Lr~I and its lanthanide homologue Lu~I, exhibiting good agreement with recent theoretical and experimental results.
    A single reference calculation is performed, allowing for substitutions from the core within a sufficiently large active set to effectively capture the influence of the core on the valence shells, improving upon previous multiconfigurational calculations.
    An additional calculation utilising a multireference set is performed to account for static correlation effects which contribute to the wavefunction.
    Reported energies for the two selected transitions are 
    20716$\pm\SI{550}{\centi \metre^{-1}}$ and 28587$\pm\SI{650}{\centi \metre^{-1}}$ for 
$7\!s^2 8s~^{2} \! {S}_{1\!/\!2}$ $\rightarrow$ $7\!s^2 7\!p ~^{2} \! {P}^{o}_{1\!/\!2 }$ 
    and
    $7\!s^2  7\!d ~^{2} \! {D}_{3\!/\!2 }$ $\rightarrow$ $7\!s^2  7\!p  ~^{2} \! {P}^{o}_{1\!/\!2 }$, respectively.
\end{abstract}

\maketitle
\section{Introduction}
\label{s:Introduction}
Much interest in the actinides has been garnered recently relating to the search for the 'island of stability' \cite{Block_Recent_2021}. 
The nuclides of the heaviest elements are generally stabilised by nuclear shell effects leading to lifetimes that allow for experimental investigation, although the actual extent of the island remains unknown \cite{Smits2023}.
Thus, the knowledge of the fundamental nuclear properties as well the influence of the nuclear shell effects in these nuclei is important to understand the limits of matter.
Atomic theory can assist in this research by predicting suitable atomic levels in the heaviest elements, which are unveiled using laser spectroscopic techniques. 
Detailed laser spectroscopy results may be combined with theoretical information on magnetic fields of the electronic shells, electric field gradients or field shift parameters. This allows the determination of nuclear moments or changes in the nuclear size, respectively.

Such experimental and theoretical collaborations have provided fruitful results for Nobelium (No, $Z=102$) \cite{Laatiaoui2016}, Fermium (Fm, $Z=100$) \cite{Warbinek2024} and for the first ionisation potentials of the heavy actinides \cite{Sato_First_2018}.
There has been a recent focus of attention on neutral lawrencium (Lr~I, $Z=103$), which concludes the actinide series of elements with active $5\!f$ orbitals. While the first ionisation potential was measured by surface ionisation, laser spectroscopic results remain pending.
In addition, before the more precise hyperfine splittings of Lr~I can be measured, the 'broad' energy level structure must first be verified \cite{Block_Recent_2021}.

Various challenges arise from studying Lr~I, for experimentalists and theorists alike.
Relativistic effects, as well as those of quantum electrodynamics (QED) 
are known to influence its atomic structure \cite{Koziol_QED_2018, Fritzsche_Lowlying_2007, Kahl_Initio_2021}. 
Comparison between experiment and theory could lead to an examination of the accuracy of multiconfigurational Dirac-Hartree-Fock (MCDHF) based methods at predicting atomic properties at high $Z$.

The challenges of measuring the atomic properties of lawrencium experimentally are exacerbated by the low production 
cross section of about 400\,nb for the isotope $^{255}$Lr in the fusion reaction of a $^{48}$Ca beam impinging on a $^{209}$Bi target \cite{Gaggeler1989}. With a primary beam of 0.8 particle \textmu A, typical target thicknesses of 0.3 mg/cm$^2$, and a separation efficiency of 30\,\%, this leads to production rates of about $0.4$ atoms \SI{}{\second^{-1}}. 
High~$Z$ elements can be produced at the Separator for Heavy Ion reaction Products (SHIP) at GSI Helmholtzzentrum für Schwerionenforschung in Darmstadt \cite{Munzenberg1979, Block2022} and are investigated by laser spectroscopy using the RAdiation-Detected Resonance Ionisation Spectroscopy (RADRIS) setup \cite{Lautenschlager2016, Warbinek_Advancing_2022}.  

RADRIS, as it stands, can detect transitions from the atomic ground state 
of Lr~I ($7\!s^2 7\!p ~^{2} \! P^{o}_{1\!/\!2}$) with transition energies between \SI{20000}{} and \SI{30000}{\centi\metre^{-1}}, and with transition rates larger than $10^7$ \invsec.
Experimental attempts have been made to measure transitions in Lr~I, with priority given to those with the highest transition rates \cite{Warbinek_Private_2023}, but still the uncertainty of the theoretical predictions requires a very large region of several hundred \si{\centi\metre^{-1}} to be scanned.

While the central scientific motivation for measuring Lr~I is driven by an interest in its atomic structure and the subsequent extraction of nuclear properties, various other applications can arise from studying the actinides, such as those in nuclear medicine. Targeted alpha therapy is a particularly exciting area of nuclear medicine in which $\alpha $-emitting radionuclides are attached to targeting vectors, which can be used to treat various diseases \cite{Nelson_Targeted_2020}.

The ground state of Lr~I is predicted to have an opposite parity relative to its lanthanide homologue, neutral lutetium (Lu~I) \cite{Desclaux_Relativistic_1980}, which is attributed to strong relativistic effects. The chemistry of Lr~I and its similarity to other elements was investigated by Xu and Pyykk\"o \cite{Xu_Chemistry_2016}. 
From the experimental side, \citet{Sato_Measurement_2015} measured the ionisation potential of Lr~I, and \citet{Kwarsick_Assessment_2021} concluded with the second ionisation potential. \citet{Sato_First_2018} experimentally verified that the $5f$ shell is fully filled at No and confirmed that the actinide series ends with Lr.

Previous calculations of Lr~I were conducted using a variety of theoretical methods, including the relativistic coupled cluster approach by \citet{Eliav_Transition_1995}, and Fock-space coupled-cluster (FSCC) by \citet{Borschevsky_Transition_2007}. A calculation of Lr~I and other elements using the combination of the configuration interaction with the linearised single-double coupled cluster method (CI + all order) was reported by \citet{Dzuba_Atomic_2014}. 
Furthermore, \citet{Kahl_Initio_2021} applied relativistic coupled-cluster method with single, double, and perturbative triple excitations (RCCSD(T)) 
alongside configuration interaction combined with many-body perturbation theory (CI+MBPT).
Moreover, \citet{Guo2024} documented the electron affinity and ionisation potential of Lr~I using the relativistic coupled cluster method.

In addition, the multireference configuration interaction model was used to provide energy levels and spectroscopic properties of the Lr$^{+}$ 
ion \cite{Ramanantoanina_Electronic_2022,Ramanantoanina_Lr_Mobilities_2023}.
Energy levels for Lr-like Rf$^{+}$ were reported by ~\citet{Ramanantoanina_Rf_2021}.
Previous MCDHF calculations of Lr~I were presented by \citet{Wijesundera_Relativistic_1995}, \citet{Zou_Resonance_2002} and \citet{Fritzsche_Lowlying_2007}. 

In this paper, Lr~I transitions of experimental interest are investigated using the MCDHF method implemented in the \grasp\ package \cite{FroeseFischer_GRASP2018_2019}, with minor modifications to ensure sufficient energy convergence and to include configurations with principal quantum numbers $n>15 $.

\section{Theory}
\label{s:Theory} 
To obtain the energy levels of a many-electron system, the eigenvalue problem can be utilised:
\begin{equation}
        \hat{H} \Psi = E \Psi.
\end{equation} 
Here $\Psi $ is the atomic state function (ASF) and the Hamiltonian $\hat{H}$ represents the many-electron Dirac-Coulomb (DC) Hamiltonian defined as
\begin{align}
    \hat{H}_{DC} &= \sum_{i=1}^{N} \left( c \boldsymbol{\alpha }_{i} \cdot \boldsymbol{p}_{i} + c^2 \left( \beta_{i} -I \right) + V_{\rm nuc} (r_{i}) \right)  \nonumber   \\ &+ \sum_{j>i=1}^{N} \frac{1}{r_{ij}} ,
    \label{eq:DC} 
\end{align}
where $r_{ij}$ represents the distance between electrons $i$ and $j$, $N$ is the total number of electrons in the system, $c$ is the speed of light, $\boldsymbol{\alpha }$ and $\beta $ are the Dirac matrices and $V_{\rm nuc} \left( r \right) $ is the nuclear potential arising from the two-parameter Fermi distribution function~\cite{Parpia_Fermi_1992}; atomic units ($\hbar = e = m_{e} = 1$) are used throughout.

In the standard formulation of the MCDHF theory, the function $\Psi  $ for a given state, is represented as a linear combination of symmetry-adapted configuration state functions (CSFs),
\begin{equation}
        \Psi \left( \Gamma \pi JM  \right) = \sum_{i=1}^{N_{\text{CSF}}} c_{i} \Phi \left( \gamma_{i} \pi JM  \right) .  
        \label{eq:mchfeq} 
\end{equation} 
Here $J$ is the total electronic angular momentum, $M$ is its projection, $\pi$ is the parity,
$\gamma_{i}$ denotes the set of orbital occupancies and complete coupling tree of angular quantum numbers unambiguously specifying the i-th CSF,
and $\Gamma $ is the identifying label, which contains all the other necessary information to uniquely describe the state function.
Configuration mixing coefficients $c_{i}$ are obtained through diagonalisation of the Hamiltonian matrix.

The relativistic configuration interaction (RCI) method is used to calculate configuration mixing coefficients $c_{i}$ without altering the orbital shapes. Corrections to the DC Hamiltonian such as the transverse photon interaction (which simplifies to the Breit interaction in the low-frequency limit), and leading-order QED corrections are included using RCI. 

After obtaining a set of atomic state functions, transition rates for an electric dipole transition between two atomic states $\Psi_{i}  $ and $\Psi_f $ can be calculated using the reduced transition matrix element,
\begin{equation}
    M_{if} = \braket{\Psi_i  | | D^{\left( 1 \right) } | |  \Psi_f },
\end{equation} 
where $D^{\left( 1 \right)}$ is the electric dipole operator in relativistic form \cite{Grant_Gauge_1974, Grant_Relativistic_2007}. 

\section{Models}
\label{s:Models}
Dirac-Hartree-Fock (DHF) wavefunctions of Lr I involve only a single CSF for each total angular momentum $J$ and parity $\pi$, representing the simplest approximation of the atomic system. However, DHF calculations are limited because they do not fully account for electron-electron repulsion, as a result of its mean field approximation. To better account for the repulsion between electrons, additional CSFs are systematically included into the wavefunction expansion. 
The electron correlation energy is the energy due to unaccounted electron-electron repulsion and is defined as the difference between DHF calculations and verified experimental results from the NIST ASD~\cite{Kramida_NIST_1999}, which are assumed to be exact for the purposes of this study. 
The CSFs accounting for the repulsion of the electrons are built from an active set (AS) of orbitals.
The AS
is systematically increased to include additional correlation orbitals as layers $\mathcal{L}_i$; a layer is defined as a set of orbitals that includes at maximum one of each orbital angular momentum symmetry, $\ell \le 4 $ or $\ell = \left\{ s,p,d,f,g \right\} $.
An additional layer increases the AS such that the AS becomes a union of the DHF spectroscopic orbitals and the layers, represented as $\text{AS}: \text{DHF} \cup \mathcal{L}_1 \cup \mathcal{L}_2 \cup \mathcal{L}_3 $ for an AS involving three layers.

The targeted states in Lr~I are the ground state $7\!s^2 7\!p ~^2 \! P^o_{1\!/\!2 }$ and the excited levels $7\!s^2 8s ~^{2} \! S_{1\!/\!2 }$, $7\!s^2 6d ~^{2} \! D_{3\!/\!2 , 5\!/\!2}$ and $7\!s^2 7\!d ~^{2} \! D_{3\!/\!2 , 5\!/\!2}$. The $7\!s^2 7\!p ~^2 \! P^o_{3\!/\!2 }$ level is also included to ensure that both the $7p_{1\!/\!2}$ and $7p_{3\!/\!2}$ orbitals are optimised. 
The results of previous calculations suggest that energies and rates of transitions from these excited states to the ground state are within the requirements of
RADRIS~\cite{Warbinek_Advancing_2022, Kahl_Initio_2021}.
To ensure computational feasibility, the number of CSFs was chosen to balance the accuracy of the wavefunction with the limitations of the computational hardware.

In quantum chemistry, electron correlation is typically divided into two categories, static correlation and dynamic correlation. Static correlation arises in atomic systems when levels are nearly-degenerate in the DHF energies, whereas dynamic correlation arises from the dynamic motions of electrons within a correlated system \cite{FroeseFischer_Computational_1997}.
Static correlation is resolved in MCDHF by 
employing a multireference (MR) set \cite{Li2020}, 
whilst dynamic correlation is more challenging and is resolved by 
systematically adding CSFs built from increasing the active sets of orbitals.
We perform a set of single reference (SR) calculations to include core correlation contributions to the wavefunction.
A separate set of calculations are performed to account for unincluded static correlation.
By including both types of these calculations, all major types of electron correlation are accounted for. 

Several factors indicate that the interelectronic interaction involving the 
$5d $ and $5f $ core subshells contributes to the energy separations.
In a heavy and neutral system, the screening effect of the inner shells diminishes the nuclear charge's influence more strongly, enhancing the core electrons influence upon the energy separations.
Additionally, the unpaired electron in Lr~I strongly polarises the core.

The calculation is divided up into four major stages or models. With each sequential model, additional correlation effects are included and the calculation is expected to become more accurate. 

\textit{Model One}: This stage of the calculations involves optimising the spectroscopic orbitals. The spectroscopic orbitals are optimised using DHF, whereby no electron substitutions are allowed from the reference set to ensure each orbital has the correct number of nodes, or number of regions where the wavefunction changes sign.
The targeted even parity levels are $7\!s^2 8s ~^{2} \! S_{1\!/\!2 }$, $7\!s^2  6d  ~^{2} \! D_{3\!/\!2 , 5\!/\!2}$, $7\!s^2 7\!d ~^{2} \! D_{3\!/\!2 , 5\!/\!2}$ and the targeted odd parity levels are $7\!s^2 7\!p ~^{2} \! P^{o}_{1\!/\!2 , 3\!/\!2}$. The $7\!s^2 6d ~^{2} \! D_{3\!/\!2 , 5\!/\!2}$ levels were not included in Model Four due to computational limitations.

In Lu~I, analogous levels were targeted. The targeted even parity levels are $6s^2 7\!s ~^{2} \! S_{1\!/\!2 }$, $6s^2  5\!d  ~^{2} \! D_{3\!/\!2 , 5\!/\!2}$ and the odd parity levels are $6s^2 6p ~^{2} \! P^{o}_{1\!/\!2 , 3\!/\!2}$. As the $6s^2 6d ~^{2} \! D_{3\!/\!2 , 5\!/\!2}$ levels are relatively high-lying in energy, these could not be optimised correctly, and were not included in the calculation.

\textit{Model Two}: The next stage of the calculations involves generating and optimising the correlation orbitals.
The strategy suggested by \citet{Papoulia_Coulomb_2019} was employed.
The MCDHF method as implemented in \grasp\ is unable to optimise all correlation orbitals simultaneously, requiring the user to adopt the layer-by-layer optimisation strategy~\cite{Schiffmann2020}.
Correlation orbitals are generated by allowing single and double (SD) substitutions from the valence shells $6d, 7\!s, 7\!p, 7\!d$ and $8s $ ($5d, 6s, 6p $ and $7\!s $) in Lr~I (Lu~I).
SD substitutions are chosen to account for dynamic correlation by accurately modelling electron-electron repulsions between two electrons.

The electric dipole transitions between the targeted states in Lr~I only involve valence electron jumps. While optimising the correlation orbitals, core substitutions were not allowed to ensure the correlation orbitals better model electron correlation in the valence orbitals and are in a close radial proximity to the valence as a result.
The AS is systematically increased up to layer ten in Lr~I with orbitals $\mathcal{L}_{10} = \left\{ 17s,16p,16d,14f,13g \right\} $.

\textit{Model Three}: The third stage of the calculation utilises the relativistic configuration interaction program \textsc{rci} as implemented in \grasp\ \cite{FroeseFischer_GRASP2018_2019}. SD electron substitutions from the previously defined valence orbitals and the $\left\{ 6s, 6p \right\} $ $\left( \left\{ 5s, 5p \right\}  \right) $ subshells in Lr~I (Lu~I) were allowed to expand the CSF basis set and account for core correlation effects. Further relativistic effects, such as those arising from the Breit interaction, as well as QED effects are added as corrections to the wavefunction \cite{Jonsson_Introduction_2022}.

\begin{figure}[h]
    \centering
    \includegraphics[width=.8\linewidth]{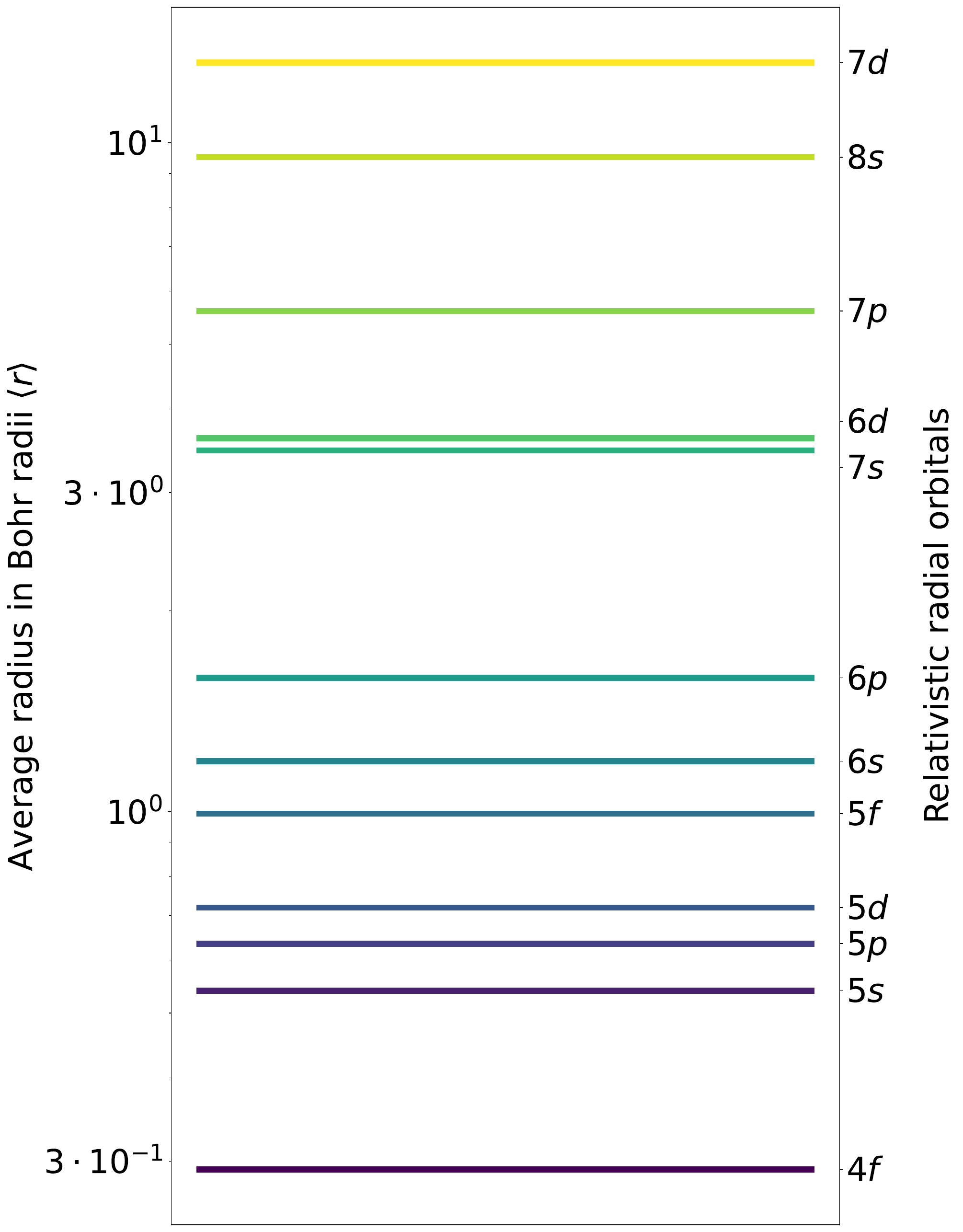}
    \caption{The average radii in Bohr radii of the spectroscopic relativistic radial orbitals in Lr I. Only radial orbitals with total angular momentum $j=l+1/2$ are included for clarity. The subshells are coloured based upon their average distance from the nucleus. }
    \label{fig:radius}
\end{figure}

Fig.~\ref{fig:radius} shows the average radius of the spectroscopic subshells in Lr~I. The correlation orbitals are optimised to be in close proximity to the valence orbitals. Furthermore, the $\left\{ 6s,6p  \right\} $ subshells are close to the valence, suggesting these would have the greatest effect on the energy separations.

\textit{Model Four}: In the fourth and largest model, \textsc{rci} is run to improve the wavefunction representation by employing a larger basis set. Further relativistic effects such as Breit, as well as QED effects are included as a correction to the wavefunction at this stage.
SD electron substitutions are allowed in Lr~I (Lu~I) from the $\left\{ 6s, 6p \right\} $ $\left( \left\{ 5s, 5p \right\}  \right) $ subshells. In addition, single and restricted double (SrD) substitutions are allowed from the $\left\{ 5d, 5f \right\} $ $\left( \left\{ 4d,4f \right\}  \right) $ subshells \cite{Bieron_Completeactivespace_2009}.
The SrD substitutions are a restriction to SD where in total two electrons are substituted however at maximum only one electron may be substituted from any of the $\left\{ 5d,5f  \right\} $ $\left( \left\{ 4d,4f  \right\}   \right) $ subshells.  
The $7\!s^2 6d ~^{2} \! D_{3\!/\!2 , 5\!/\!2}$ levels could not be included in Model Four due to computational limitations.
The SrD method is designed to model core-valence correlation effects without considering costly core-core correlation effects, which are expected to have a negligible impact on energy separations.

To address the effects of static correlation, a multireference (MR) set is created which includes configurations with largest contribution to the wavefunction.
To obtain the configurations with the greatest contribution, a small test calculation was conducted in SR with two layers using Model Two. Configurations from this calculation were sorted by mixing coefficient and added to the MR set until the cumulative mixing coefficients reached 95\% of the wavefunction.
Calculations were performed with the MR set similarly to the main SR calculation, with the exception that the MR calculation does not use Model Four due to computational limitations.

\textit{MR Model One}: 
For the first stage of the MR calculation, the Lr~I MR set includes the odd parity configuration $7\!s 6d 7\!p $ in addition to the $7\!s^2 7\!p $ configuration and the even parity configuration $7\!s 7\!p^2$ in addition to the $7\!s^2 8s $, $7\!s^2  6d  $ and $7\!s^2 7\!d $ configurations.
The spectroscopic orbitals were optimised using DHF.
The targeted even parity levels were $7\!s^2 8s ~^{2} \! S_{1\!/\!2 }$, $7\!s^2  6d  ~^{2} \! D_{3\!/\!2 , 5\!/\!2}$, $7\!s^2 7\!d ~^{2} \! D_{3\!/\!2 , 5\!/\!2}$ and the targeted odd parity levels are $7\!s^2 7\!p ~^{2} \! P^{o}_{1\!/\!2 , 3\!/\!2}$.

The calculation is repeated for lutetium, the Lu~I MR set includes the odd parity configuration $6s 5d 6p$ in addition to the $6s^2 6p $ configuration, and the even parity configurations $6s 5d^2 $, $6s 6p^2 $ are included in addition to the $6s^2 7\!s $, $6s^2  5\!d  $ configurations.
For Lu~I, analogous levels were targeted. The even parity targeted levels are $6s^2 7\!s ~^{2} \! S_{1\!/\!2 }$, $6s^2  5\!d  ~^{2} \! D_{3\!/\!2 , 5\!/\!2}$ and the odd parity levels are $6s^2 6p ~^{2} \! P^{o}_{1\!/\!2 , 3\!/\!2}$. 

\textit{MR Model Two:}
In the second stage of the MR calculation, correlation orbitals are generated and optimised.
Correlation orbitals are generated by allowing single and double (SD) substitutions from the valence shells $6d, 7\!s, 7\!p, 7\!d$ and $8s $ ($5d, 6s, 6p $ and $7\!s $) from the Lr~I (Lu~I) MR set.

\textit{MR Model Three:}
In the third and final stage of the MR calculation, SD substitutions from the previously defined valence shells and electron substitutions from the $\left\{ 6s, 6p \right\} $ $\left( \left\{ 5s, 5p \right\}  \right) $ subshells were allowed to further expand the CSF basis and consider core contributions.
A calculation is performed on the expanded CSF basis set using RCI, further relativistic and QED effects are included as a correction to the wavefunction.

\section{Results and discussion}
\label{s:Results}
\begin{figure*}
    \centering
    \includegraphics[width=\linewidth]{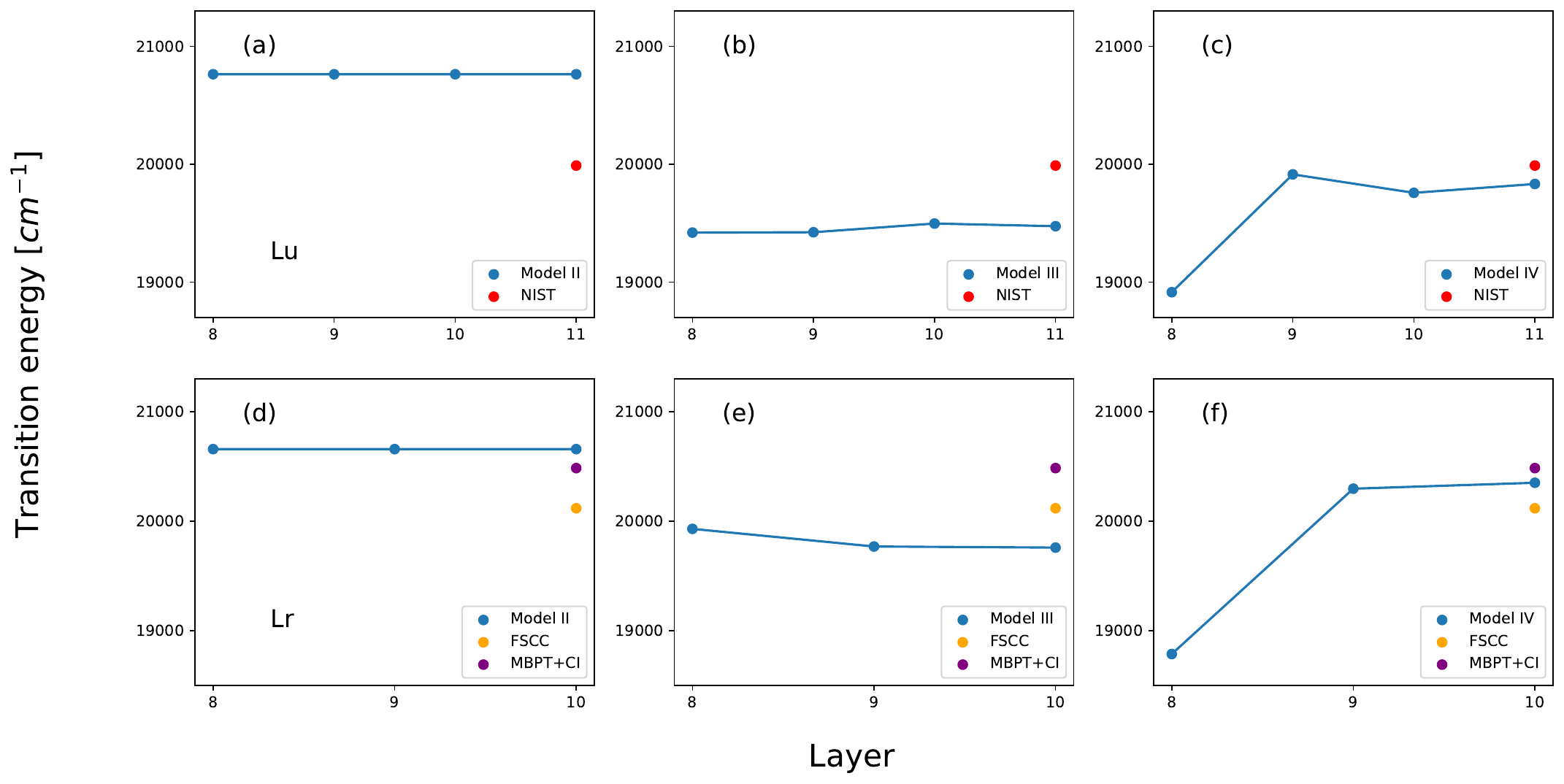} 
    \caption{MCDHF calculations of the transition energy for the last four computational layers of Lr~I (Lu~I), as a function of computational layer for the transition of
    $7\!s^2 8s ~^{2} \! S_{1\!/\!2 }$ $\rightarrow$ $7\!s^2 7\!p ~^{2} \! P^{o}_{1\!/\!2}$ 
    ($6s^2 7\!s ~^{2} \! S_{1\!/\!2 }$ $\rightarrow$ $6s^2 6p ~^{2} \! P^{o}_{1\!/\!2}$).
    The upper three subfigures (a)-(c) present the results for Lu~I while the lower three subfigures (d)-(f) present the results for Lr~I. The subfigures from left to right show how the calculations evolve as the computational model becomes larger and more core correlation is considered.
        The subfigures on the left (a), (d) show the transition energies for Model Two, when only valence effects are considered. The middle subfigures (b), (e) show the transition energies for Model Three. 
        The subfigures to the right (c), (f) show the transition energies for Model Four. 
        The calculations of Lr I are compared to the results of the previous theoretical calculations using FSCC~\cite{Borschevsky_Transition_2007} and CI+MBPT \cite{Kahl_Initio_2021}, while the calculations of Lu I are compared to experimental values from NIST \cite{Kramida_NIST_1999}.  \label{fig:all_conv}}
\end{figure*}

Figs. \ref{fig:all_conv} (a) - (f) show how the calculations converge as the AS is systematically increased.
The total energy of the system decreases as the AS increases, however the even and odd parities will decrease at a different rate, agreeing at convergence \cite{Schiffmann_Electronic_2021}.
Allowing substitutions from the core magnifies oscillations due to the large amounts of correlation associated with the core, and increasing the size of the AS is required to compensate this.

\begin{table*}[t]
    \centering
    \caption{The calculated energy levels of Lr~I and Lu~I for different computational models, details on the different models are discussed in section \ref{s:Models}. The results are compared to NIST \cite{Kramida_NIST_1999} where available, the values in the 'Total' column are obtained by considering the effects of both static and dynamic correlation. }
    \label{tab:main}
    \begin{ruledtabular}
\begin{tabular}{ccccccccc}
    \multicolumn{3}{c}{Levels}          & \multicolumn{6}{c}{Energy (\si{\centi \metre^{-1}})}                                \\ 
    Element & Configuration & Term                      & Model Two & Model Three & MR Model Three & Model Four & Total & NIST  \\ \hline
    Lu      & $6s^2 5d$     & $~^{2} \! D_{3\!/\!2 }$   & 0         & 0           & 0              & 0          & 0     & 0     \\
            & $6s^2 5d$     & $~^{2} \! D_{5\!/\!2 }$   & 590       & 1695        & 1619           & 1778       & 1702  & 1993  \\
            & $6s^2 6p$     & $~^{2} \! P^o_{1\!/\!2 }$ & 1334      & 3548        & 2358           & 3864       & 3864  & 4136  \\
            & $6s^2 6p$     & $~^{2} \! P^o_{3\!/\!2 }$ & 3765      & 6866        & 5686           & 7220       & 7230  & 7476  \\
            & $6s^2 7s$     & $~^{2} \! S_{1\!/\!2 }$   & 21353     & 23023       & 23523          & 23717      & 24217 & 24125 \\ \hline
    Lr      & $7s^2 7p$     & $~^{2} \! P^o_{1\!/\!2 }$ & 0         & 0           & 0              & 0          & 0     & -     \\
            & $7s^2 6d$     & $~^{2} \! D_{3\!/\!2 }$   & 4907      & 3145        & 4853           & -          & -     & -     \\
            & $7s^2 6d$     & $~^{2} \! D_{5\!/\!2 }$   & 7106      & 6047        & 7763           & -          & -     & -     \\
            & $7s^2 7p$     & $~^{2} \! P^o_{3\!/\!2 }$ & 8133      & 8283        & 8540           & 8223       & 8480  & -     \\
            & $7s^2 8s$     & $~^{2} \! S_{1\!/\!2 }$   & 20658     & 19930       & 22003          & 20351      & 20716 & -     \\
            & $7s^2 7d$     & $~^{2} \! D_{3\!/\!2 }$   & 28543     & 27700       & 29922          & 28073      & 28587 & -     \\
            & $7s^2 7d$     & $~^{2} \! D_{5\!/\!2 }$   & 28694     & 27949       & 30166          & 28277      & 28786 & -
\end{tabular}
\end{ruledtabular}

\end{table*} 

Model Four is designed to capture all leading dynamic correlation contributions, while the MR Model Three calculations are designed to capture static correlation effects.
Unfortunately, it is not currently computationally feasible to perform calculations of Lr~I that fully account for both static and dynamic correlation.
To account for both types of electron correlation, a method is proposed to calculate the change in energy separations as a result of accounting for static electron correlation.

Table \ref{tab:main} presents the results of the SR and MR calculations of Lu~I and Lr~I. These indicate that using Model Four is essential for achieving good agreement between the even and odd parity energy levels. However, Model Four is computationally infeasible with the MR.

As a result, the contributions due to the MR are computed by considering only the change in energy between SR Model Three and MR Model Three for that level and the ground state level of the same parity.

Our method calculates the total energy by taking the value of the energy for a level in Model Four and subtracting the energy difference due to the inclusion of the MR set, which is derived from the energy differences of levels with the same parity,
\begin{align}
    E_{\text{Total}} &= E_{\text{Dynamic}} + E_{\text{Static}} \nonumber \\ 
    &= E_{\text{Model Four}} + \Delta E_{\text{MR}}.
    \label{eqs:tot} 
\end{align}
For the case of the $7s^2 8\!s ~^{2} \! S_{1\!/\!2 }$ energy level in Lr~I (Table~\ref{tab:main}), the change in energy due to the inclusion of the MR set, $\Delta E_{\text{MR}}$, is computed as $\left( 22003 - 4853 \right) - \left( 19930 - 3145 \right)  = \SI{365}{\centi \metre^{-1}}$, which can be then combined with the Model Four result to give the total energy separation.

For all excited levels in Lr~I, $\Delta E_{\text{MR}}$ is positive, meaning that including the MR leads to an increase in parity-relative energy separations. Similarly, the effect of accounting for additional core correlation appears to be an increase in parity-relative energy separations.

Notably, after applying this method the $6s^2 7\!s ~^{2} \! S_{1\!/\!2 }$ level in the Total column in Table~\ref{tab:main} is remarkably close to the NIST value. However, the $6s^2  5d  ~^{2} \! D_{5\!/\!2 }$ level becomes further from the NIST value. Possible explanations for this could include an unbalanced MR between the ground and excited energy levels.
The $7\!s^2  6d  ~^{2} \! D_{3\!/\!2, 5\!/\!2 }$ levels could not be included in Model Four due to computational limitations.

\begin{table*}
    \caption{The transition rates for Lu~I and Lr~I are showcased for different computational models, the results of Lu~I are compared to NIST\cite{Kramida_NIST_1999} and Lr~I is compared to CI+MBPT\cite{Kahl_Initio_2021}. The Lu~I transition rates of Models Two and Three had poor energy separations and were scaled with the factor $\lambda_{\text{Calculated}}^3 / \lambda_{\text{NIST}}^3$ where $\lambda $ is the transition wavelength to allow comparison between models. 
    \label{tab:Lu_trans}}
\begin{ruledtabular}
\begin{tabular}{cccccccccc}
            & \multicolumn{2}{c}{Upper levels} & \multicolumn{2}{c}{Lower levels} & \multicolumn{5}{c}{Einstein A coefficient (\si{\second^{-1}})} \\
    Element & Conf      & Term                    & Conf      & Term                      & Model Two      & Model Three    & Model Four     & MR Model Three & NIST/CI+MBPT   \\ \hline
    \rule{0pt}{0.35cm}
    Lu      & $6s^2 7s$ & $~^{2} \! S_{1\!/\!2 }$ & $6s^2 6p$ & $~^{2} \! P^o_{1\!/\!2 }$ & \num{3.14E+07} & \num{4.10E+07} & \num{4.21E+07} & \num{3.21E+07} & \num{3.20E+07} \\
            & $6s^2 7s$ & $~^{2} \! S_{1\!/\!2 }$ & $6s^2 6p$ & $~^{2} \! P^o_{3\!/\!2 }$ & \num{5.23E+07} & \num{6.94E+07} & \num{7.25E+07} & \num{5.45E+07} & \num{4.9E+07}  \\ \hline
    \rule{0pt}{0.35cm}
    Lr      & $7s^2 8s$ & $~^{2} \! S_{1\!/\!2 }$ & $7s^2 7p$ & $~^{2} \! P^o_{1\!/\!2 }$ & \num{3.13E07}  & \num{3.41E07}  & \num{3.30E07}  & \num{3.38E07}  & \num{3.31E07}  \\
            & $7s^2 7d$ & $~^{2} \! D_{3\!/\!2 }$ & $7s^2 7p$ & $~^{2} \! P^o_{1\!/\!2 }$ & \num{5.03E07}  & \num{4.91E07}  & \num{4.61E07}  & \num{3.95E07}  & \num{6.14E07}  \\
            & $7s^2 8s$ & $~^{2} \! S_{1\!/\!2 }$ & $7s^2 7p$ & $~^{2} \! P^o_{3\!/\!2 }$ & \num{3.31E07}  & \num{3.34E07}  & \num{3.39E07}  & \num{2.88E07}  & \num{3.57E07}  \\
            & $7s^2 7d$ & $~^{2} \! D_{3\!/\!2 }$ & $7s^2 7p$ & $~^{2} \! P^o_{3\!/\!2 }$ & \num{9.57E06}  & \num{9.80E06}  & \num{9.75E06}  & \num{6.94E06}  & \num{1.21E07}  \\
            & $7s^2 7d$ & $~^{2} \! D_{5\!/\!2 }$ & $7s^2 7p$ & $~^{2} \! P^o_{3\!/\!2 }$ & \num{3.68E07}  & \num{5.19E07}  & \num{5.19E07}  & \num{3.92E07}  & \num{5.39E07}
\end{tabular}
\end{ruledtabular}


\end{table*}
Table \ref{tab:Lu_trans} shows the calculated transition rates in Lu~I and Lr~I across the different models, and compares these to both experiment and previous theory.
Interestingly, the Lu~I calculations are close to the experimental values for Model Two and the MR Model Three, but not accurate for Models Three and Four, this shows the importance of including the MR when calculating transition rates.

Model Two can mimic the MR calculation by allowing SD substitutions between the valence shells, allowing many different configurations to be created.
However, the same is not true for Models Three and Four.

In Models Three and Four, where core substitutions are allowed, if one electron moves from a valence shell to another to make a new valence configuration, only one core electron can be substituted due to the SD restriction. This single core substitution is inadequate for accurately representing electron-electron interactions in the core, resulting in the new configuration having a relatively minimal contribution to the system's total energy.
This may be resolved by either allowing single, double and triple (SDT) electron substitutions or by conducting a calculation with an MR set.

\begin{table*}[t]
    \caption{The calculated energy levels of experimental significance of Lr~I in comparison to previous theory. \label{tab:Lr_prev}}
    \begin{ruledtabular}
\begin{tabular}{ccccccccc}   
    \multicolumn{2}{c}{Levels}          & \multicolumn{7}{c}{Energy (\si{\centi \metre^{-1}})  }                                \\ 
    Configuration  & Term    & This work & CI + MBPT \cite{Kahl_Initio_2021} & RCCSD(T) \cite{Kahl_Initio_2021} & FSCC \cite{Borschevsky_Transition_2007} & CI + all order \cite{Dzuba_Atomic_2014} & MCDHF \cite{Fritzsche_Lowlying_2007} & MCDHF \cite{Zou_Resonance_2002}\\ \hline
     $7s^2 7p $ & $~^{2} \! P^o_{1\!/\!2 }$ & 0     & 0     & 0     & 0     & 0     & 0     & 0    \\
     $7s^2 7p $ & $~^{2} \! P^o_{3\!/\!2 }$ & 8480  & 8606  & 8677  & 8413  & 8495  & 8138  & 7807 \\
     $7s^2 8s $ & $~^{2} \! S_{1\!/\!2 }$ & 20736 & 20485 & 20533 & 20118 & 20253 & 20405 & -    \\
     $7s^2 7d $ & $~^{2} \! D_{3\!/\!2 }$ & 28607 & 28580 & -     & 28118 & -     & -     & -    \\
     $7s^2 7d $ & $~^{2} \! D_{5\!/\!2 }$ & 28806 & 28725 & -     & 28385 & -     & -     & -   \\  
\end{tabular}
\end{ruledtabular}
 
\end{table*}

Given the difference in transition rates between Models Three and Four are small, it is assumed that the difference between MR Model Three and a hypothetical MR Model Four would also be negligible. Therefore, results from experiment and previous theory are compared to MR Model Three.
As Models Two and Three in Lr~I yield energy separations that deviate significantly from the NIST values, these transition rates are adjusted by the factor $\lambda^3_{\text{Calculated}} / \lambda^3_{\text{NIST}}$ to align with the experimental energy separations and to allow comparison between models.

The Lu~I and Lr~I transition rates are compared to NIST and CI+MBPT \cite{Kahl_Initio_2021}, respectively.
The Lu~I transition rates show excellent agreement with experiment with the differences between NIST being 0.3 and 11\% for the two calculated transitions in Table~\ref{tab:Lu_trans}.
The calculated transition rates of Lr~I also show good agreement with previous theory, especially for transitions originating from the upper level of $7\!s^2  8s  ~^{2} \! S_{1\!/\!2}$, which show deviations of 2.1\% and 23\% compared to CI+MBPT.
Although the transition rates from the $7\!s^2  7\!d  ~^{2} \! D_{3\!/\!2, 5\!/\!2 }$ upper levels are approximately 35\% larger than the values in MR Model Three, this discrepancy is within the uncertainity of 40\% estimated by \citet{Kahl_Initio_2021}.

Table~\ref{tab:Lr_prev} presents a comparison between our calculated results and previous theoretical values, with our results showing the closest agreement with the recent CI+MBPT and RCCSD(T) values reported by \citet{Kahl_Initio_2021}.
Previous MCDHF calculations have reported the $7\!s^2 7\!p ~^2 \! P^o_{3\!/\!2 } -  7\!s^2 7\!p ~^2 \! P^o_{1\!/\!2 }$ energy to be lower than our value. This may be attributed to the limited inclusion of electron correlation in those studies.

Our calculations improve upon those of \citet{Zou_Resonance_2002} and \citet{Fritzsche_Lowlying_2007} which did not include the effects of the $g$ correlation orbitals and core contributions from the $5d $ orbitals, respectively.
The $g$ correlation orbitals are important for describing the actinides since these orbitals have a direct dipole interaction with the $5\!f $ core orbitals \cite{Fritzsche_Lowlying_2007}.
Furthermore, due to the computational constraints that existed at the time, convergence is not demonstrated in either work as only two layers were included in the calculations.

\subsection{Uncertainty}
\label{ss:er}
Assuming the uncertainties are independent and uncorrelated, the total energy uncertainty can be calculated as the square root of the sum of the squares of the dynamic and static energy uncertainties \cite{Taylor1997}.

The uncertainty in energy is calculated for the transitions from the upper levels $7\!s^2 8s ~^{2} \! S_{1\!/\!2 }$, $7\!s^2 7d ~^{2} \! D_{3\!/\!2 }$ to the ground state $7\!s^2 7p ~^{2} \! P_{1\!/\!2 }^o$. 
The static uncertainty is taken to be the difference between the MR Model Three energy relative to the lowest energy level of the same parity and the Model Three energy relative the lowest energy level of the same parity in Table~\ref{tab:main}. This is calculated to be $\SI{365}{\centi \metre^{-1}}$ and $\SI{514}{\centi \metre^{-1}}$ for the $7\!s^2 8s ~^{2} \! S_{1\!/\!2 }$ and $7\!s^2 7d ~^{2} \! D_{3\!/\!2 }$ levels, respectively.

The dynamic uncertainty is taken to be the difference between the NIST ASD and the Lu~I Model Four values. The corresponding level of the $7\!s^2 8s ~^{2} \! S_{1\!/\!2 }$ level in lutetium is the $6s^2 7s ~^{2} \! S_{1\!/\!2 }$ level, the dynamic uncertainty for this level is determined to be $\SI{408}{\centi \metre^{-1}}$. The $7\!s^2 7\!d ~^{2} \! D_{3\!/\!2 }$ level has no lutetium equivalent due to computational limitations, therefore it is assumed the dynamic uncertainty for this level is the same as for level $7\!s^2 8s ~^{2} \! S_{1\!/\!2 }$.

The total uncertainty can be obtained by taking the square root of the sum of the squares of the two uncertainties, this is then rounded to the nearest $\SI{50}{\centi \metre^{-1}}$. The uncertainty of the $7\!s^2 8s ~^{2} \! S_{1\!/\!2 }$ level in Lr~I is reported as $\SI{547}{\centi \metre^{-1}}$, rounded to $\SI{550}{\centi \metre^{-1}}$ while the uncertainty of $7\!s^2  7\!d  ~^{2} \! D_{3\!/\!2 }$ level is $\SI{656}{\centi \metre^{-1}}$, rounded to $\SI{650}{\centi \metre^{-1}}$.

\section{Conclusion}
\label{s:Conclusion} 
We report calculated values of experimentally significant energy levels and corresponding transition rates of Lr~I and its lighter homologue Lu~I, obtained using the MCDHF method as implemented in \grasp.
Calculations were performed with various computational models with and without a multireference set to capture different types of electron correlation.

The reported transitions energies are 20716$\pm \SI{550}{\centi \metre^{-1}}$ for the atomic transition $7\!s^2 8s~^{2} \! {S}_{1\!/\!2}$ $\rightarrow$ $7\!s^2 7\!p ~^{2} \! {P^{o}}_{1\!/\!2 }$ and 28587$\pm \SI{650}{\centi \metre^{-1}}$ for the $7\!s^2  7\!d ~^{2} \! {D}_{3\!/\!2 }$ $\rightarrow$ $7\!s^2  7\!p  ~^{2} \! {P}^{o}_{1\!/\!2 }$ atomic transition.
This work improves upon previous MCDHF calculations by \citet{Zou_Resonance_2002} and \citet{Fritzsche_Lowlying_2007} by demonstrating convergence and by considering the effects of the $g$ correlation orbitals and core contributions from the $5d $ orbitals.
Calculated energy levels and transition rates of Lu~I and Lr~I exhibited good agreement with experiment and previous theory, respectively.
Our work on Lr~I appears to be in closest agreement with the theoretical calculations performed using CI+MBPT and RCCSD(T) \cite{Kahl_Initio_2021}. 
The demonstrated agreement within different atomic theory frameworks will help the effort of experimental level search in Lr~I
as it clearly restrains the region which needs to be investigated.
The uncertainties of the calculated energy levels, arising from limited computational resources, were quantified and remain large compared to the discrepancies observed in individual recent theoretical calculations.

Future calculations may include a multireference model variationally to consider both static and dynamic correlation within a large active set. To accomplish this, new methods could be utilised such as independently optimised orbital sets \cite{Li_Independently_2022} or machine learning \cite{Bilous_DeepLearning_2023} to reduce the computational load.

\section*{Acknowledgments}
The authors would like to acknowledge and thank Jessica Warbinek for her helpful discussions.
This project has received funding from the European Union’s Horizon 2020 research and innovation programme under grant agreement No 861198–LISA–H2020-MSCA-ITN-2019.
J. G. thanks the Swedish Research Council for the individual starting grant with contract number 2020-05467.
P. J. acknowledges support from the Swedish research council under contract 2023-05367.
These calculations were performed in part on the HPC cluster "Draco" provided by Friedrich-Schiller Universit\"at Jena.

\bibliographystyle{apsrev4-2-author-truncate}
\bibliography{bib_big.bib}    
\end{document}